\documentclass[twocolumn,showpacs,preprintnumbers,amsmath,amssymb]{revtex4}

\usepackage{graphicx}
\usepackage{dcolumn}
\usepackage{bm}

\begin{document}

\preprint{Brown-NIST Team}

\title{Fate of the Peak Effect in a Type-II
Superconductor: Multicriticality in the Bragg-Glass Transition\\
}

\author{S. R. Park$^1$, S. M. Choi$^2$, D. C. Dender$^3$, J. W. Lynn$^3$, and X. S. Ling$^{1\dagger}$}
\address{$^1$Department of Physics, Brown University, Providence, Rhode Island 02912 USA\\
$^2$Korea Advanced Institute of Science and Technology, Taejon, South Korea\\
$^3$NIST Center for Neutron Research, Gaithersburg, Maryland
20899 USA\\}

\date{\today}

\begin{abstract}
We have used small-angle-neutron-scattering (SANS) and ac
magnetic susceptibility
 to investigate the global magnetic field $H$ vs temperature
 $T$ phase diagram of a single crystal Nb in which a first-order transition of
 Bragg-glass melting (disordering), a peak effect, and surface superconductivity
 are all observable.   It was found that the disappearance of the peak effect is directly related
 to a multicritical behavior in the Bragg-glass transition.
 Four characteristic phase boundary lines have been identified on the $H-T$ plane:
 a first-order line at high fields, a mean-field-like continuous transition line
 at low fields, and two continuous transition line associated with the onset of
 surface and bulk superconductivity.  All four lines are found to meet at a multicritical point.
\end{abstract}

\maketitle

An outstanding question concerning the Abrikosov vortex state of
type-II superconductors is whether a genuine order-disorder
transition can still occur in vortex matter even though true
crystalline order cannot be attained due to random pinning by
impurities \cite{larkin}.  There are convincing theoretical
arguments \cite{nat,gl,hwa,fis} and numerical evidence \cite{huse}
suggesting that, instead of a true vortex crystal, a novel Bragg
glass phase with quasi-long-range order can exist in bulk samples
with weak random pinning; hence a true order-disorder
transition can occur when the topological order of the Bragg
glass is destroyed, by thermal fluctuations and/or random
pinning.

However, it is still controversial as to whether a Bragg glass
melting (or disordering) transition is the underlying mechanism
of the well-known anomaly of ``peak effect'' in weak-pinning
type-II superconductors \cite{menon}. Recent neutron scattering experiments on
Nb \cite{ling}, and V$_3$Si \cite{gapud}, as well as STM studies
of 2H-NbSe$_2$ \cite{kes} all suggested a disordering transition
at the peak effect, and the phase transition appears to be first
order \cite{ling}. However, it has been reported that some
samples of similar quality, e.g., having only weak
pointlike pinning centers, do not show a peak effect, nor a
disordering phase transition \cite{forgan}. This raises an
obvious, but intriguing, question:  Is the fate of the peak
effect, i.e. appearing or disappearing, related to a
multicritical behavior in the phase transition into the Bragg glass? In
this Letter, we report the first direct evidence that the
disappearance of the peak effect is related to a multicritical
point on the Bragg-glass phase boundary.

Our experiment was carried out using the 30m SANS instruments NG7
and NG3 at the NIST Center for Neutron Research on a Nb single
crystal (99.998$\%$ in purity) in which both the peak effect and
the first-order Bragg-glass melting (disordering) transition were
observed at the same temperatures \cite{ling}. The sample has a
zero-field $T_c$ = 9.16 K, and an estimated Ginzburg-Landau
parameter $\kappa_1$(0) = 2.0. The mean wavelength of the
incident neutron beam was $\lambda$ = 6.0 {\AA} and the
wavelength spread 11 $\%$ (FWHM). The experimental configuration
is shown in the inset of Fig.1 (a). A cadmium mask was used such
that only the central portion of the sample was exposed to the
incoming neutron beam. The scattered neutrons were captured by a
2D detector of 128x128 pixels (the pixel size is 0.5 cm by 0.5 cm)
15.3 m away from the sample. The dc magnetic field was applied in
the direction of the incoming neutron beam using a horizontal
superconducting magnet. A coil was wound on the sample to allow
{\it in situ} ac magnetic susceptibility measurements.

Fig.1(a) shows the SANS data at $H$= 3.0 kOe. The Gaussian width
data are obtained from fitting the measured (Bragg) intensity vs.
azimuthal angle to six Gaussian peaks evenly spaced 60$^o$
apart.  It is clear that the azimuthal widths, a measure of
orientational disorder in the vortex array, are strongly history
dependent. Supercooling and superheating effects are observed for
field-cooling (FC) and field-cooled-warming (FCW) paths,
respectively.  As reported previously \cite{ling},
 the disordered phase at $T > T_p$ and the ordered phase at $T < T_p$
 are of
 their respective thermodynamic
 ground states.
 The abrupt change in the
structure factor $S(q)$ at the peak effect $T_p$ depicts a
symmetry-breaking phase transition from a vortex matter with
short-range order to a Bragg glass with quasi-long range order
\cite{ling}.  The phase transition is first order as evidenced by
the strong thermal hysteresis in $S(q)$. Compared to that at
higher fields, the metastability region for $H$ = 3.0 kOe is
smaller but still pronounced.

We found that the thermal hysteresis of $S(q)$ observed in SANS is
strongly field dependent, and the metastability region disappears
completely at low fields.  Fig.1 (b) shows the azimuthal width
data for $H$ = 2.0 kOe.  For comparison, the real part $\chi'(T)$
of the ac magnetic susceptibility is also shown in Fig.1(b).  The
dip in $\chi'(T)$ is a well-established signature of the peak
effect \cite{ling1,ishida,shi}.  The history dependence of Bragg-peak width
is detectable only within 0.1 K of the peak-effect temperature
$T_p$.

A similar trend is observable in the history dependence of the
radial widths of the Bragg peaks, as shown in Fig.2, obtained by
fitting a single Gaussian function to the $q$-dependence of the
SANS intensity.  At 3.0 kOe, there is a pronounced thermal
hysteresis in the radial widths.  At 2.0 kOe, however, the
hysteresis is barely discernable.  At an even lower field of 1.0
kOe (data not shown), the thermal hysteresis in $S(q)$ is
undetectable. At $H$ = 1.0 kOe, a very sharp peak effect (the
onset-to-end width = 40 mK) is still present.  Thus we believe the
phase transition at 1.0 kOe is still first-order but the
metastability region is too narrow to be resolved in SANS (the
temperature resolution in SANS $\approx$ 50 mK). Nevertheless,
the diminishing hysteresis in the Bragg-glass transition in the
low-field regime suggests that the phase transition is becoming
continuous and mean-field-like.

The fact that the transition into the Bragg glass is first order
at high fields, but continuous (mean-field-like) at low fields
strongly suggests the existence of a multicritical point on the
phase boundary bordering the Bragg-glass on the $H-T$ phase
diagram. We show below that this multicritical behavior is directly
related to the appearance and the disappearance of the peak
effect.

Fig.3 shows a three-dimensional plot of the $\chi'(T)$ as a
function of temperature and magnetic field in the field range of
0 - 5.12 kOe. At high fields, there is a pronounced peak effect,
a characteristic dip in $\chi'(T)$. With decreasing field, the
peak effect becomes narrower and smaller. For $H <$ 0.8 kOe,
there is only a single kink in $\chi'(T)$ corresponding to the
mean-field transition $H_{c2}(T)$. According to the
$\chi'(T)$ data in Fig.3, there is no reentrant peak effect at
low fields, in contrast to that in 2H-NbSe$_2$ \cite{ghosh} and
YBCO \cite{ling2}. The peak effect simply vanishes here.

At higher temperatures above the peak-effect temperature $T_p(H)$
(or $H_p(T)$, used interchangeably), there is a smooth step in
$\chi'(T)$.  This step, $T_{c3}(H)$ (or $H_{c3}(T)$), defined in
Fig.1 (b), is the onset of surface superconductivity. The separation
between $T_p$ and $T_{c3}$ grows larger with increasing
magnetic field.  Upon cooling, below $T_{c3}(H)$ and towards
$T_p(H)$, the screening effect in $4\pi\chi'(T)$ increases
gradually. Nevertheless, a less well-defined characteristic
temperature $T_{c2}(H)$ can be identified to mark the onset of bulk 
superconductivity [see Fig. 1(b) for definition]. Note that the notation
$T_{c2}(H)$ is used for $H >$ 0.8 kOe, while $H_{c2}(T)$ is used for $H <$ 0.8 kOe.

The results of Fig.3 are summarized in a new phase diagram of
Bragg glass superconductivity in Nb as shown in Fig.4. The measured 
ratio of $H_{c3}/H_{c2}$ at low temperatures is about 1.60,
slightly smaller than the expected value of 1.695 by Saint-James
and de Gennes \cite{saint}.  This is likely due to the nonideal
cylinder surface being not exactly parallel to the field . The
crossing of $H_{c3}(T)$ and $H_{c2}(T)$ lines below $T_c$ was
observed previously \cite{cruz,finn}, and has been interpreted as due
to a depressed BCS gap function near the surface \cite{hu}.

The most striking aspect of Fig.4 is that all four lines,
$H_p(T)$, $T_{c2}(H)$, $H_{c2}(T)$, and $H_{c3}(T)$, meet at a 
multicritical point (MCP). To determine the nature of a MCP,
one needs to know how many of these lines are second-order 
phase transitions. In the theory of Saint-James and de Gennes \cite{saint},
$T_{c3}(H)$ is a continuous phase transition. For $H <$ 0.8 kOe, 
the linear temperature dependence of $H_{c2}(T)$ follows the expected 
behavior of a Ginzburg-Landau mean-field transition line \cite{helfand}. 
This is a line of continuous phase transitions from the normal state directly into
an ordered Abrikosov Bragg-glass phase. For $H  >$ 0.8 kOe, the
peak effect $T_p(H)$ traces out a line of first-order
transitions. Across this line, the thermal hysteresis in the
structural factor $S(q)$ of the vortex matter can be observed,
especially striking at high fields.  

The nature of the $T_{c2}(H)$ line is less clear. The vortex matter
is liquidlike structurally in the shaded part of the phase diagram.
Whether this disordered vortex matter is a distict thermodynamic phase
from the normal state is still being debated (see \cite{menon} and
references therein). If $T_{c2}(H)$ is a true second-order
phase transition, e.g., as in a vortex glass transition \cite{fisher},
the MCP in Fig. 4 would appear to be a tricritical point. On the other hand,
we found that the measured slopes of the four phase boundaries near the 
MCP cannot satisfy the requirements \cite{yip} imposed on a tricritical
point by thermodynamics, but are consistent with those for a bicritical point.
This leads to an important conclusion that only one of the two lines,
$T_{c2}(H)$ or $H_{c3}(T)$, can be related to the MCP.

For $T_{c2}(H)$ to be a second-order line for the bicritical point,
its slope has to be larger (in magnitude) than that of $H_{c2}(T)$, such
that the thermodynamics rule \cite{wheeler} that no phase can
occupy more than 180$^o$ of the phase space around a MCP is satisfied.
However, this is not observed in our data as shown in Fig. 4. For
the $H_{c3}(T)$ line to be the relevant one, the ratio of specific-heat
jump at $H_{c2}(T)$ over that at $H_{c3}(T)$ should be 43.6. While
this large ratio is consistent with the existing specific-heat data
on Nb \cite{serin}, presently there are no reliable specific-heat
data near a crossing point of $H_{c2}(T)$ and $H_{c3}(T)$ to allow
us to make a quantitative comparison.

We should point out that a similar critical point has also been
observed in platelet geometries such as MgB$_2$ \cite{welp,pissas}
crystals for which $H_{c3}(T)$ is not expected to play a role in
the critical point. In high-$T_c$ YBCO, a disappearance of the
first-order transition was also observed in the low-field regime,
and was interpreted as a critical end point \cite{kwok}. If our
critical point in Fig. 4 is also interpreted as a critical end
point, $T_{c2}(H)$ would not be a true phase transition, and the meeting of 
$H_{c3}(T)$ at the MCP would be purely coincidental. 

In summary, we found that, in a Nb crystal in which a peak effect
in ac magnetic susceptibility and a first-order melting
(disordering) transition in SANS were found to coincide
previously, both effects disappear at a low field.  It is
suggested that the appearance or absence of a peak effect in a
type-II superconductor may be directly correlated with a
multicritical point (MCP) on the Bragg glass phase boundary. The 
existence of a MCP, at which the peak effect vanishes, suggests
that the origin of the peak effect is related to a second-order 
phase transition at a higher temperature. In the sample studied,
it appears that the MCP may be related to surface superconductivity.

Two of us (S.R.P. and X.S.L.) are indebted to Prof. P.H.
Kes and his group in Leiden University for kind hospitality and
stimulating discussions. During the course of this work, we benefited from
numerous discussions with Professors. A. Houghton, J.M. Kosterlitz,
M.C. Marchetti, D.R. Nelson, R.A. Pelcovits, H.H. Wen, and S.C.
Ying. X.S.L. wishes to thank Prof. Cees Dekker and the 
Delft University of Technology (the Netherlands) for hosting
his sabbatical visit in academic year 2002-2003, and the
Guggenheim Foundation for financial support.
This work was supported by the National Science Foundation under
Grants No. DMR-0102746 and No. DMR-9986442, and the Galkin Fund
(S.R.P.).      \\
\\
\ \\
{\bf Figure Captions:}
\\
Figure 1: (a) Temperature and history dependence of azimuthal
widths of the (1,-1) diffraction peak at $H_{dc}$ = 3.0 kOe. The
widths are obtained by Gaussian fittings. The dashed line is the
peak of the peak effect $T_p$ at this magnetic field based on ac
magnetic susceptibility measurements.  Inset: Experimental
configuration. (b) Temperature and history dependence of the
azimuthal widths of (1,-1) diffraction peak at $H_{dc}$ = 2.0
kOe.  The ac susceptibility data are also shown for reference.
Definitions of $T_p(H)$, $T_{c2}(H)$, and $T_{c3}(H)$ (see below) are shown.

Figure 2: Temperature and history dependence of the radial widths
of the diffraction peaks at (a) $H_{dc}$ = 3.0 kOe, and (b)
$H_{dc}$ = 2.0 kOe.

Figure 3: Three-dimensional (3D) magnetic field and temperature
dependence of the real part of the ac susceptibility
$4\pi\chi'(T)$. $H_{dc}||H_{ac}$. Note that two values of ac
fields were used in the measurements. For $H_{dc} <$3.0 kOe,
$H_{ac}$ = 1.7 Oe and $f$ = 1.0 kHz and for $H_{dc} >$ 3.0 kOe,
$H_{ac}$ =  7.0 Oe and $f$ = 1.0 kHz. The solid and dashed lines
are guides to eyes. For the ac fields used, $T_p$ is independent
of the ac field amplitude \cite{ling1}.

Figure 4: The phase diagram of a weak-pinning Nb crystal for the
$H_{dc} ||$ $<111>$ crystallographic direction. The upper solid
circles (the crosses correspond to measurements using $H_{ac}$ =
1.7 Oe, see Fig.3 captions) are the peak of the peak effect and
the first-order transition line; the lower ones are the mean-field
transition. The open diamonds (two sets are for two values of ac
fields, see Fig.3) are $H_{c3}$. The multicritical point is
indicated by the large filled circle. The $H_{c1}$ data (triangle) are
estimated from the first penetration in the ac susceptibility
data.  All lines are hand drawn as guides for eyes.

$^{\dagger}$Author to whom correspondence should be addressed.
Electronic address: xsling@brown.edu.

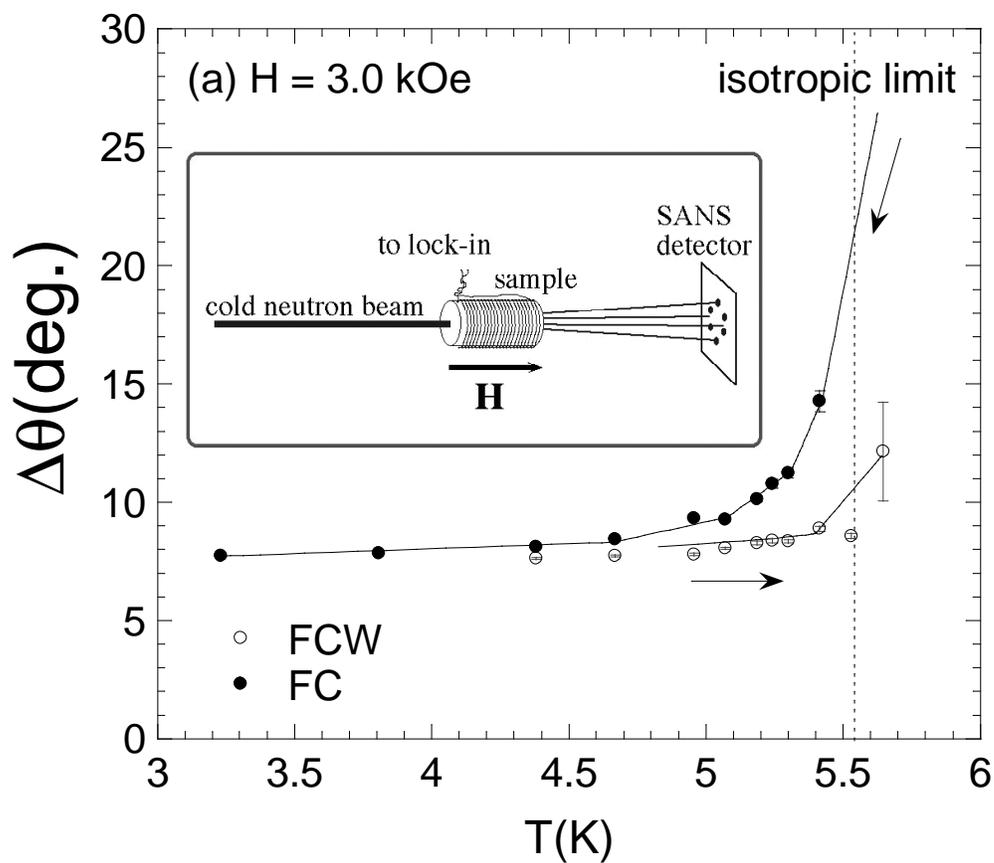
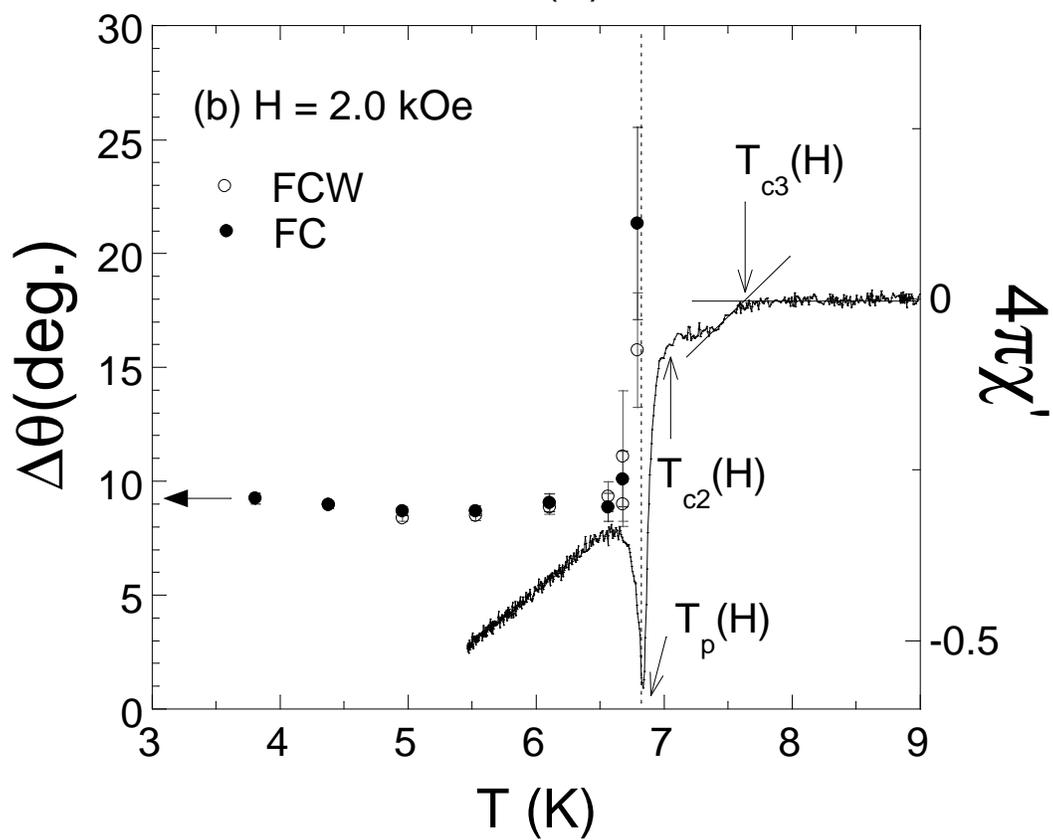
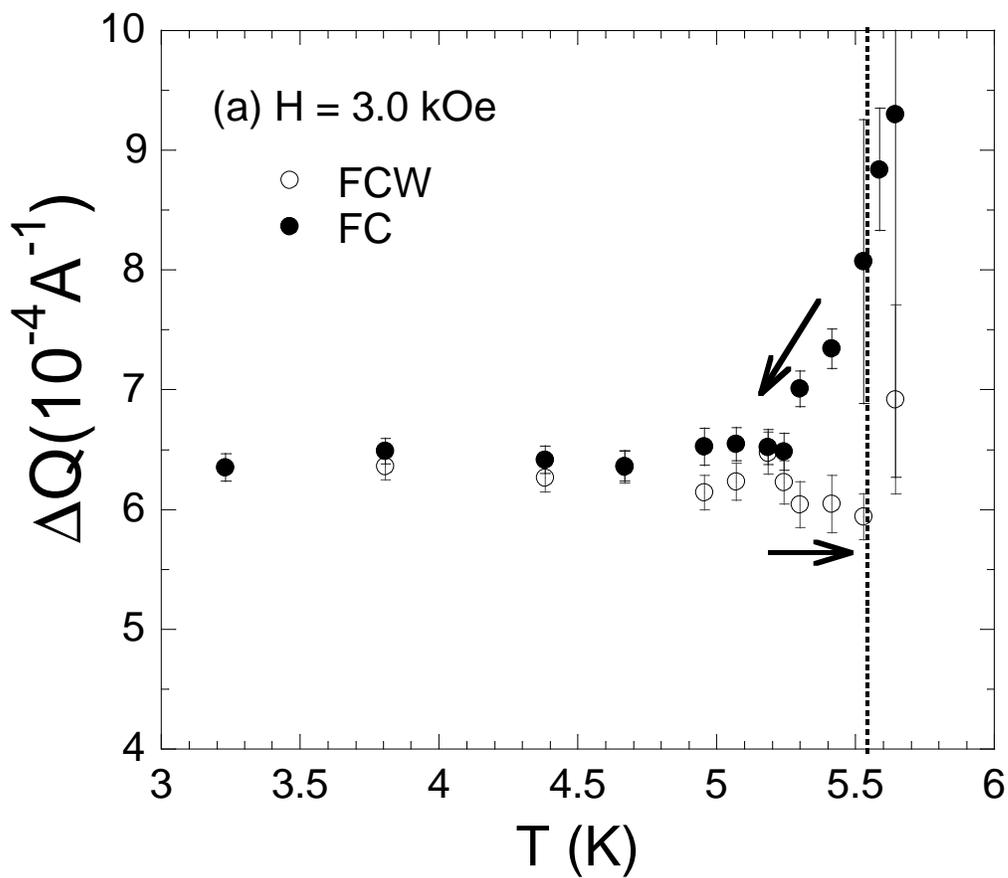
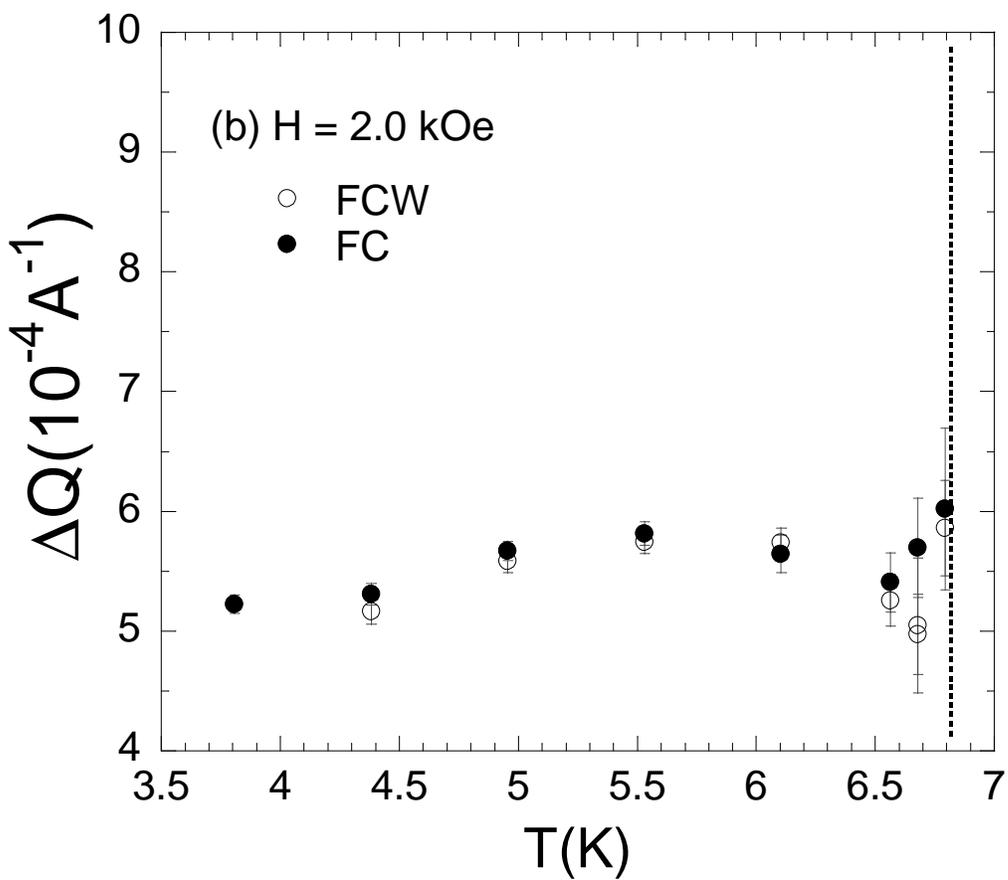
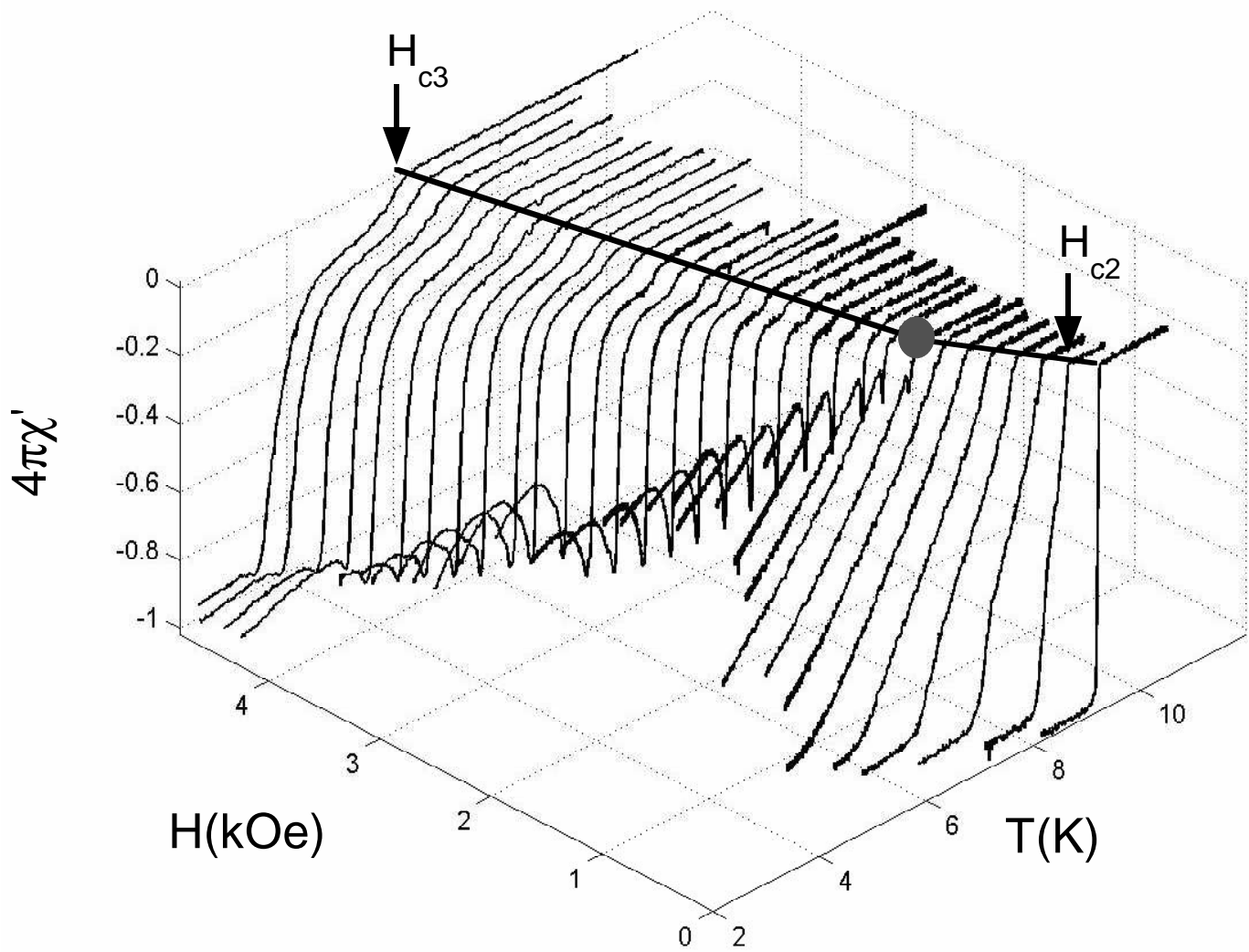
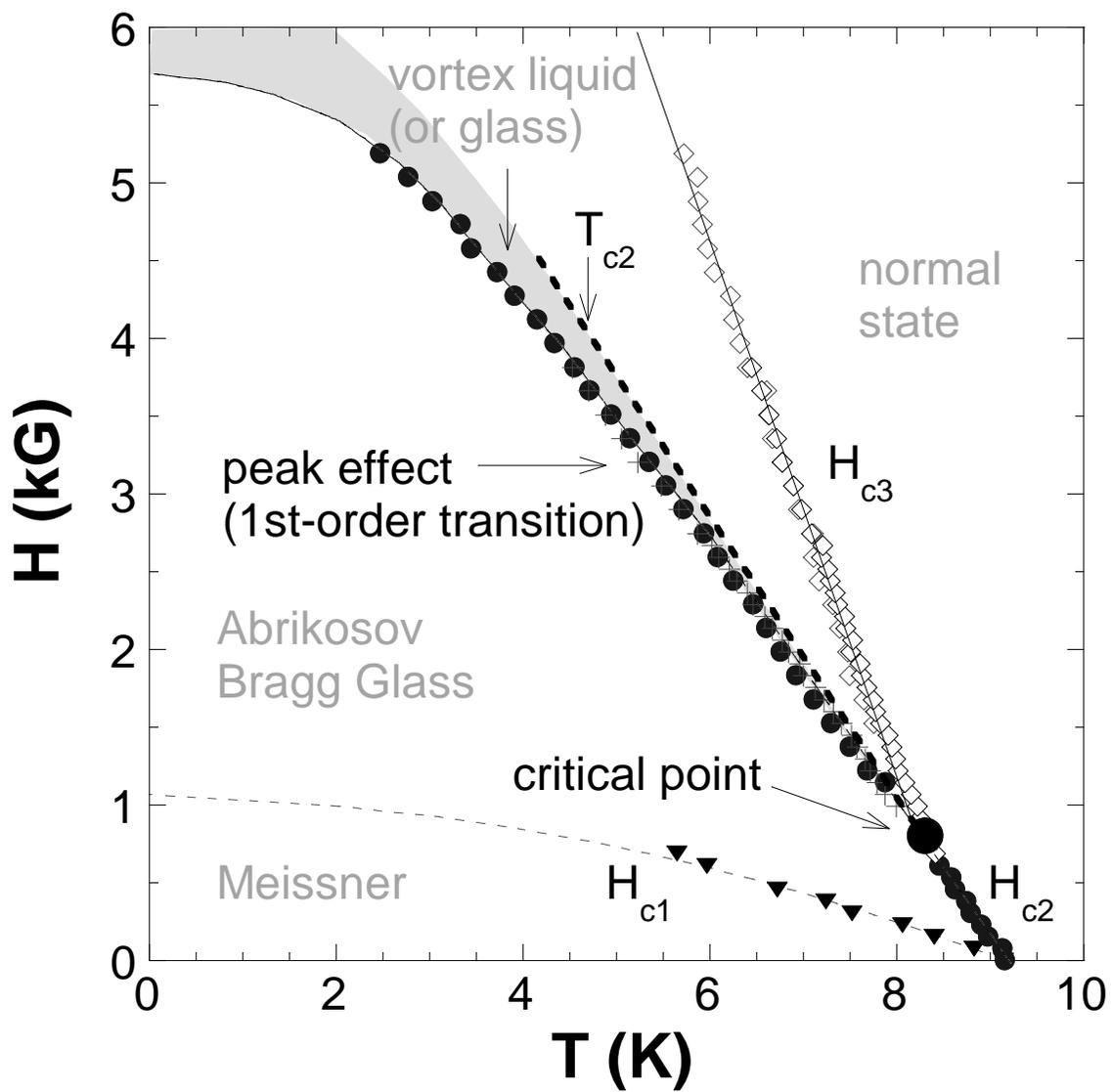
\end{document}